\documentclass{article}
\usepackage{spconf,amsmath,graphicx}
\usepackage{siunitx}
\usepackage{mystyle}
\usepackage{multirow}
\usepackage{booktabs}
\usepackage{xcolor}
\usepackage{hyperref}
\usepackage{textcomp}
\usepackage[utf8]{inputenc}

\usepackage{tikz}
\usepackage{tikz-3dplot}
\usetikzlibrary{arrows.meta}
\usetikzlibrary{positioning}

\makeatletter
\tikzoption{canvas is plane}[]{\@setOxy#1}
\def\@setOxy O(#1,#2,#3)x(#4,#5,#6)y(#7,#8,#9)%
{\def\tikz@plane@origin{\pgfpointxyz{#1}{#2}{#3}}%
\def\tikz@plane@x{\pgfpointxyz{#4}{#5}{#6}}%
\def\tikz@plane@y{\pgfpointxyz{#7}{#8}{#9}}%
\tikz@canvas@is@plane
}
\makeatother 
 
\tdplotsetmaincoords{70}{30}

\title{A Hierarchical Subspace Model for Language-Attuned Acoustic Unit Discovery}
\name{Bolaji Yusuf$^{\star \dagger}$ \qquad Lucas Ondel$^{\dagger}$ \qquad {Lukáš Burget}$^{\dagger}$ \qquad Jan Černocký$^\dagger$ \qquad {Murat Saraçlar}$^{\star}$}

  \address{$^{\star}$ Boğaziçi University, Department of Electrical and Electronics Engineering, Istanbul, Turkey \\
      $^{\dagger}$ Brno University of Technology, Faculty of Information Technology, Speech@FIT, Czechia
      \thanks{The work was supported by Czech National Science Foundation (GACR) project "NEUREM3" No. 19-26934X, Czech Ministry of Education, Youth and Sports project No. LTAIN19087 "Multi-linguality in speech technologies", and by European Union’s Horizon 2020 project No. 870930 - WELCOME.}
      }
      
\begin{document}
\ninept
\maketitle
\begin{abstract}
    In this work, we propose a hierarchical subspace model for acoustic unit discovery. In this approach, we frame the task as one of learning embeddings on a low-dimensional \emph{phonetic} subspace, and simultaneously specify the subspace itself as an embedding on a \emph{hyper}-subspace. We train the hyper-subspace on a set of transcribed languages and transfer it to the target language. In the target language, we infer both the language and unit embeddings in an unsupervised manner, and in so doing, we simultaneously learn a subspace of units specific to that language and the units that dwell on it. We conduct our experiments on TIMIT and two low-resource languages: Mboshi and Yoruba. Results show that our model outperforms major acoustic unit discovery techniques, both in terms of clustering quality and segmentation accuracy.
% This work introduces a hierarchical subspace model for the task of acoustic unit discovery. We represent acoustic units as embeddings on a low-dimensional manifold; we represent the manifold's own parameters as embeddings on a hyper-subspace encoding phonetic variability across languages. The model is trained with a two-stage procedure: first, the hyper-subspace parameters are estimated from several transcribed source languages, then the model discovers units in a target language while the hyper subspace parameters remain fixed. We conduct our experiments on TIMIT and two low-resource languages: Mboshi and Yoruba. Results show that our model outperforms major acoustic unit discovery techniques, both in terms of clustering quality and segmentation accuracy.
%
\end{abstract}

\section{Introduction}
\label{sec:intro}
Current machine learning approaches for speech processing rely on large collections of annotated audio recordings. In contrast, infants learn to speak long before they are able to read and write.
%  They naturally structure speech into phones, syllables and words to compose a sentence.
 Augmenting machines with similar capability would have a great impact. First, it would drastically reduce the cost of data annotation, therefore allowing speech technologies to be extended to low-resource languages. Second, by the proposed ``reverse-engineering'' approach to cognitive science \cite{dupoux2018cognitive}, it would pave the way to a better understanding of human learning.% as well as more practical artificial intelligence. 

In this paper, we focus on the task of Acoustic Unit Discovery (AUD). This task consists of discovering an inventory of phone-like units---denoted ``acoustic units''---from a set of untranscribed recordings. This is a simplified model of a language acquisition where we consider learning phonetics rather than the complete structure of speech (phones, syllables, words, ...).

The AUD task has been the subject of numerous publications~\cite{versteegh2015zero, dunbar2017zero, dunbar2019zero}. Nowadays, two major approaches are widely used: (i) neural-network-based models which typically use auto-encoder structure with a discretization layer \cite{chorowski2019unsupervised,Baevski2020vq-wav2vec,van2017neural} (ii) non-parametric Bayesian generative-based models which can be seen as infinite mixtures time series models \cite{lee2012nonparametric,ondel2016variational, chen2017multilingual}, or hybrids of both as in~\cite{glarner2018full}.

This work follows the Bayesian paradigm and is a direct extension of \cite{Ondel2019shmm}, where the target language's acoustic units parameters are forced to lie on a language-independent phonetic subspace that is estimated from several transcribed languages.

We propose a subspace that is adapted to the target-language in an unsupervised fashion. We achieve this by learning a language-independent \emph{hyper}-subspace from transcribed data in other languages, and a low-dimensional embedding vector for the target (low-resource) language. The hyper-subspace is a set of matrices which can be thought of as subspace ``templates" and the embedding determines how these templates are combined for the target language. Thus we have \emph{hierarchical} structure in which the lower level constrains units and the higher level constrains subspaces.

%  We propose to adapt this subspace to the target language by learning a language specific embedding. This embedding defines a combination of ``template subspace'' bases which, together, form the phonetic subspace of the target language.

\section{Problem definition}
\label{sec:model}
% \begin{figure}[tp]
%     \centering
%     \includegraphics[width=0.35\textwidth,angle=90]{figures/HGSM_graph_new.pdf}
%     \caption{Graphical model of the proposed model}
%     \label{fig:my_label}
% \end{figure}
% \begin{figure}[t]
%     \centering
%     \input{figures/hshmm_graph}
%     \caption{Graphical model of the proposed model}
%     \label{fig:hshmm_graph}
% \end{figure}

% \subsection{Problem formulation}
The problem of acoustic unit discovery can be formulated as that of learning a set of $U$ discrete units with parameters $\Matrix{H} = \{\Matrix{\eta}^1, \dots, \Matrix{\eta}^U\}$ from a sequence of untranscribed acoustic features $\Matrix{X}=(\Matrix{x}_1, \dots, \Matrix{x}_N)$, as well as the assignment of frame to unit $\Matrix{z} = (z_1, \dots, z_N)$. Formally, we seek to maximize:
\begin{align}
    p(\Matrix{z}, \Matrix{H} | \Matrix{X}) &\propto p(\Matrix{X} | \Matrix{z}, \Matrix{H}) p(\Matrix{z}, \Matrix{H})
    \label{eq:bayes}.
\end{align}

As in~\cite{lee2012nonparametric,ondel2016variational}, $p(\Matrix{X} | z_n, \Matrix{H})$ is given by an HMM with parameters $\Matrix{\eta}^{z_n}$, and we further factorize the prior:
\begin{align}
    p(\Matrix{z}, \Matrix{H}) = p(\Matrix{z} | \Matrix{H}) \prod_{u=1}^{U}p(\Matrix{\eta}^u).
\end{align}
Note that the number of units $U$ is unknown and also needs to be learned for an unknown language. Prior work \cite{ondel2016variational} addresses this issue by constructing $p(\Matrix{z}|\Matrix{H})$ with a sample from a Dirichlet process~\cite{Teh2010a} with base measure $p(\Matrix{\eta})$. This leads to an infinite ``phone-loop'' model where each acoustic unit component is a 3-state left-to-right HMM. The exact relation between the Dirichlet Process and the phone-loop structure of the model is discussed at length in \cite{ondel2020thesis}. In this work we focus on the construction of the base measure and we leave the rest of the model unaltered.
% Note that the number of units $U$ is unknown and also needs to be learned for an unknown language. To get a model with an arbitrary (theoretically infinite) number of units, where the actual number of units is inferred from the data, we define the prior of the distribution of conditional priors over units $p(\Matrix{z}|\Matrix{H})$ to be a Dirichlet process~\cite{Teh2010a} with base measure $p(\Matrix{\eta})$.

The base measure $p(\Matrix{\eta})$ defines a prior probability that a sound---represented by an HMM with parameters $\Matrix{\eta}$---is an acoustic unit. Earlier works on Bayesian AUD~\cite{lee2012nonparametric, ondel2016variational, ondel2018bayesian, ondel2017bayesian} use exponential family distributions as the base measure. These distributions, while mathematically convenient since they form conjugate priors, do not incorporate any knowledge about phones. For instance, the models \textit{a priori} may consider the sound of a car engine to be as likely an acoustic unit as the ``ah" sound. Perhaps more detrimentally, the models are also likely to model other sources of variability such as speaker, emotional state, channel etc.
% The base measure $p(\Matrix{\eta})$ defines a prior probability that a sound---represented by an HMM with parameters $\Matrix{\eta}$---is an acoustic unit. Earlier works on Bayesian AUD~\cite{lee2012nonparametric, ondel2016variational, ondel2018bayesian, ondel2017bayesian} set the base measure as combinations of exponential family probability distributions to have analytically tractable posteriors over the parameters. While these are mathematically convenient, they do not incorporate any knowledge about the distribution of speech. For instance, \textit{a priori} the models consider the sound of a car engine to be as likely an acoustic unit as the ``ah" sound. Perhaps more detrimentally, the models are also likely to model other sources of variability such as speaker, channel etc.

Therefore, we utilize the generalized subspace model (GSM)~\cite{Ondel2019shmm} which provides a solution to the problem of specifying an educated base measure by defining the parameters of each unit $u$ as:
\begin{align}
    \Matrix{\eta}^u =f( \Matrix{W} \cdot \Matrix{e}^u + \Matrix{b}),
\end{align}
% correct
where $\Matrix{e}^u$ is a low-dimensional unit embedding, $\Matrix{W}$ and $\Matrix{b}$ are the subspace parameters and $f(\cdot)$ is a deterministic and differentiable function that ensures that the resulting vector $\Matrix{\eta}^u$ dwells in the HMM parameter space. $\Matrix{W}$, $\Matrix{e}^u$, and $\Matrix{b}$ are assumed to have Gaussian distributions with diagonal covariance matrices. The posteriors of $\Matrix{W}$ and $\Matrix{b}$ are estimated from other, transcribed, languages and fixed for the target language while the posteriors of $\Matrix{e}^u$ are learned in the target language. Thus, the parameters $\Matrix{H} = \{\Matrix{\eta}^1, \dots, 
\Matrix{\eta}^U\}$ are constrained to a low-dimensional manifold of the parameter space.% This model was shown to improve over its unconstrained counterparts and provides the starting point for the hierarchical subspace model proposed in this work.

\begin{figure}[t!]
    \centering
    \begin{tikzpicture}[scale=1.3, tdplot_main_coords]
        \input{figures/hyperspace}
    \end{tikzpicture}
    \caption{Illustration of a hierarchical subspace model. For each language $\lambda$, acoustic unit embeddings (encoding the parameters of a probabilistic model) are assumed to live in a language-specific subspace $\textcolor{red!70}{\Matrix{W}^{\lambda}}$ of the total parameter space. This subspace is given by a weighted sum of matrix bases $\textcolor{blue!70}{\Matrix{M}_1}, \textcolor{blue!70}{\Matrix{M}_2}, \dots$ (shared across languages) and language-specific weights $\Matrix{\alpha}^{\lambda}$: $\textcolor{red!70}{\Matrix{W}^{\lambda}} = \alpha_1^{\lambda} \textcolor{blue!70}{\Matrix{M}_1} + \alpha_2^{\lambda} \textcolor{blue!70}{\Matrix{M}_2} + \dots$.}
    \label{fig:hsubspace}
\end{figure}

\section{Hierarchical Subspace HMM}
% Hierarchical subspace model
% \subsection{Hierarchical Subspace Hidden Markov Model}
The GSM of~\cite{Ondel2019shmm} enforces an educated prior by transferring the subspace parameters ($\Matrix{W}, \Matrix{b}$) to a target language. This makes the implicit assumption that the subspace is universal i.e. that the units of all languages lie on the same manifold. We hypothesize that this is too strong an assumption, and that having a language-dependent subspace allows us to better model the units of a specific target language. However, naively training, or even fine-tuning, the subspace on the target language counters its purpose by removing the constraint on the space of units, thereby losing transferred phonetic information.

To deal with  the dilemma, we propose a hierarchical subspace model (HSM). The crux of this model is to allow language-dependent subspaces, but only as long as they lie on a manifold of the \emph{hyper-space} of subspaces, as depicted Fig. \ref{fig:hsubspace}. Formally:
\begin{align}
    \Matrix{W}^\lambda &= \Matrix{M}_0 + \sum_{k=1}^K \alpha^\lambda_k \Matrix{M_k} \label{eq:W} \\
    \Matrix{b}^\lambda &= \Matrix{m}_0 + \sum_{k=1}^K \alpha^\lambda_k \Matrix{m_k} \label{eq:b} \\
    \Matrix{\eta}^{\lambda,u} &= f(\Matrix{W}^\lambda \cdot \Matrix{e}^{\lambda, u} + \Matrix{b}^\lambda),
    \label{eq:eta}
\end{align}
where $\Matrix{W}^\lambda$ and $\Matrix{b}^\lambda$ define the subspace for language $\lambda$ and $\Matrix{\eta}^{\lambda,u}$ is the vector of parameters for unit $u$ of language $\lambda$. The unit parameters, $\Matrix{\eta}^{\lambda, u}$ are constructed from a linear combination of the columns of $\Matrix{W}^\lambda$ weighted by unit-specific embedding vectors $\Matrix{e}^{\lambda, u}$ and a bias vector $\Matrix{b}^\lambda$. Similarly, $\Matrix{W}^\lambda$ is defined by a linear combination of basis matrices $[\Matrix{M}_1, \dots, \Matrix{M}_K]$ weighted by language-specific embedding vectors $\Matrix{\alpha}^{\lambda} = [\alpha^\lambda_1, \alpha^\lambda_2, \dots, \alpha^\lambda_K]^\top$ plus bias matrix $\Matrix{M}_0$. The bias vector $\Matrix{b}^\lambda$ is similarly obtained by a linear combination of $[\Matrix{m}_1, \dots, \Matrix{m}_K]$ and $\Matrix{\alpha}^{\lambda}$ plus bias term $\Matrix{m}_0$.

We assume Gaussian priors for the random variables:
% \begin{align}
%     \alpha^\lambda_k, M_{k,ij}, e^{{\lambda},u}_i & \sim \Gaussian (0, 1).
% \end{align}
\begin{align}
    \alpha^\lambda_k & \sim \Gaussian (0, \sigma_\alpha) \\
    M_{k,ij} & \sim \Gaussian (0, \sigma_M) \\
    m_{k,i} & \sim \Gaussian (0, \sigma_m) \\
    {e}^{{\lambda},u}_i & \sim \Gaussian (0, \sigma_e),
\end{align}
with variances set to 1. Note that the posterior distribution that we seek is modified from~\eqref{eq:bayes} to:
\begin{align}
    p(\Matrix{z}, \Matrix{E}^{\lambda}, \Matrix{\alpha}^{\lambda}, \mathcal{M} | \Matrix{X}^\lambda) \propto \hspace{1mm} & p(\Matrix{X}^\lambda | \Matrix{z}, \Matrix{E}^{\lambda}, \Matrix{\alpha}^{\lambda}, \mathcal{M}) p(\Matrix{z} | \Matrix{E}^{\lambda}, \Matrix{\alpha}^{\lambda}, \mathcal{M}) \nonumber \\
    & \cdot p(\Matrix{E}^{\lambda})p(\Matrix{\alpha}^{\lambda})p(\mathcal{M})
    \label{eq:bayes_mod}
\end{align}
$\mathcal{M}=(\Matrix{M}_0, \dots, \Matrix{M}_K, \Matrix{m}_0, \dots, \Matrix{m}_K)$, $\Matrix{E}^\lambda = \{\Matrix{e}^{\lambda,1}, \dots, \Matrix{e}^{\lambda,U^\lambda}\}$ and $\Matrix{X}^\lambda$ is language-specific data. To complete the definition of our generative process, we model the likelihood of a speech segment $\Matrix{X}_s^\lambda $ given an acoustic unit $p(\Matrix{X}_s^\lambda|\Matrix{z}_s=u, \Matrix{H})$ as a 3-state left-to-right HMM with parameter vector:
% \begin{equation}
%     \Matrix{\eta}^{\lambda, u} = \begin{bmatrix}
%     \Matrix{\eta}^{\lambda, u}_{1}\\
%     \Matrix{\eta}^{\lambda, u}_{2}\\
%     \Matrix{\eta}^{\lambda, u}_{3}
%     \end{bmatrix},
% \end{equation}
\begin{equation}
    \Matrix{\eta}^{\lambda, u} = \begin{bmatrix}
    \Matrix{\eta}^{\lambda, u \top}_{1},
    \Matrix{\eta}^{\lambda, u \top}_{2},
    \Matrix{\eta}^{\lambda, u \top}_{3}
    \end{bmatrix}^\top,
\end{equation}
and each state has emission probabilities modeled as a GMM with $K=4$ Gaussian components:
% \begin{equation}
%     \Matrix{\eta}^{\lambda, u}_i = \begin{bmatrix} \Matrix{\mu}_{\lambda, u}^{i,1} \\ \vdots \\ \Matrix{\mu}_{\lambda, u}^{i,K} \\ \Vectorize(\Matrix{\Sigma}_{\lambda, u}^{i,1}) \\ \vdots \\ \Vectorize(\Matrix{\Sigma}_{\lambda, u}^{i,K}) \\ \pi_{\lambda, u}^{i,1} \\ \vdots \\ \pi_{\lambda, u}^{i,K} \end{bmatrix},
% \end{equation}
\begin{align}
    \Matrix{\eta}^{\lambda, u}_i = \big[& \Matrix{\mu}^{\lambda, u \top}_{i,1}, \dots, \Matrix{\mu}^{\lambda, u \top}_{i,K}, \Vectorize(\Matrix{\Sigma}^{\lambda, u}_{i,1})^\top, \dots,  \Vectorize(\Matrix{\Sigma}^{\lambda, u }_{i,K})^\top, \nonumber \\
    & \pi^{\lambda, u}_{i,1} \dots \pi^{\lambda, u}_{i,K } \big]^\top,
\end{align}
where $[\cdot]^\top$ is the transpose operator, $\Vectorize$ is the vectorize operator, ${\pi}_{i,j}^{\lambda, u}$, $\Matrix{\mu}_{i,j}^{\lambda, u}$ and $\Matrix{\Sigma}_{i,j}^{\lambda, u}$ are the weight, mean and covariance matrix of the $i$th HMM state and the $j$th Gaussian component of the acoustic unit $u$ in language $\lambda$. The function $f(\cdot)$ in~\eqref{eq:eta} is defined as:
\begin{align}
    \pi^{\lambda, u}_{i,j} &= \frac{\exp\{\Matrix{W}^{\lambda, i}_{\pi} \cdot \Matrix{e}^{\lambda, u} + \Matrix{b}^{\lambda, i}_{\pi}\}_j }{1 + \sum_{k=1}^{K-1}\exp\{\Matrix{W}^{\lambda, i}_{\pi} \cdot \Matrix{e}^{\lambda, u} + \Matrix{b}^{\lambda, i}_{\pi} \}_k} \label{eq:f_definition_pi} \\
    \Matrix{\Sigma}^{\lambda, u}_{i, j} &= \Diag (\exp\{ \Matrix{W}^{\lambda, i, j}_{\Sigma} \cdot \Matrix{e}^{\lambda, u} + \Matrix{b}^{\lambda, i}_{\Sigma}  \})
    \label{eq:f_definition_sigma} \\
    \Matrix{\mu}^{\lambda, u}_{i, j} &= \Matrix{\Sigma}^{\lambda, u}_{i, j} \cdot \big( \Matrix{W}^{\lambda, i, j}_{\mu} \cdot \Matrix{e}^{\lambda, u} + \Matrix{b}^{\lambda, i}_{\mu}  \big) \label{eq:f_definition_mu},
\end{align}
where $\exp$ is the element-wise exponential function and $\exp \{...\}_j$ is the $j$th element of the resulting vector. $\Matrix{W}^{\lambda, i}_{\pi}$ is the subset of rows of matrix $\Matrix{W}^{\lambda}$ assigned to the mixing weights $\Matrix{\pi}^{\lambda,\cdot}_{i}$ of the $i$th HMM state. Matrices $\Matrix{W}^{\lambda, i,j}_{\mu}$ and $\Matrix{W}^{\lambda, i,j}_{\Sigma}$ are similarly defined for the mean and covariance matrix of the $j$th Gaussian component of $i$th HMM state.

Thus we have a Hierachical Subspace Hidden Markov Model (H-SHMM). Note that the choice of HMM as the likelihood model follows previous work~\cite{lee2012nonparametric,ondel2016variational,Ondel2019shmm} and is not integral to the proposed hierarchical subspace model.
% We leave the exploration of other models such as deep generative models to future work.

\subsection{Inference in the Hierarchical Subspace HMM}
The H-SHMM training procedure follows the SHMM training~\cite{Ondel2019shmm} modified to accommodate the alterations made to the model. Given a set of $L$ languages, our goal is to compute the parameters' posterior:%we use variational Bayes to approximate the posterior over parameters:
%Given a set of $L$ languages, our goal is to maximize:
\begin{align}
    p(\{ \Matrix{z}^\lambda \}, \{ \Matrix{E}^\lambda \}, \{ \Matrix{\alpha}^\lambda \}, \mathcal{M} | \{ \Matrix{X}^\lambda \}), \quad \lambda \in \{1, \dots, L \}, \label{eq:post}
\end{align}
% \begin{align}
%     p(\Matrix{z}^1, \dots,\Matrix{z}^L, \Matrix{E}^1, \dots, \Matrix{E}^L, \Matrix{\alpha}^1,\dots, \Matrix{\alpha}^L,\nonumber \\ 
%     \mathcal{M}| \Matrix{X}^1, \dots, \Matrix{X}^L) \nonumber \\
%     = \prod_{\lambda=1}^L p(\Matrix{z}^\lambda, \Matrix{E}^\lambda, \Matrix{\alpha}^\lambda,\mathcal{M}|\Matrix{X}^\lambda), \label{eq:post}
% \end{align}
where $\Matrix{z}^\lambda, \Matrix{E}^\lambda,\Matrix{\alpha}^\lambda$ and $\Matrix{X}^\lambda$ are language-specific variables. For conciseness, we define $\Matrix{\theta}^\lambda = (\{\Matrix{\alpha}^\lambda\}, \{\Matrix{E}^\lambda\})$. Since \eqref{eq:post} is intractable, we seek an approximate posterior $q$ by maximizing the variational lower-bound $\mathcal{L}[q]$ subject to the following mean-field factorization:
\begin{align}
    q(\{ \Matrix{z}^\lambda \}, \{ \Matrix{\theta}^\lambda \}, \mathcal{M}) &= \Big[ \prod_{\lambda = 1}^L q(\Matrix{z}^\lambda) q(\Matrix{\theta}^\lambda) \Big] q(\mathcal{M}) \nonumber \\
    &= q(\{ \Matrix{z}^\lambda \} )q(\{\Matrix{\theta}^\lambda \})q(\mathcal{M}). \label{eq:meanfield}
\end{align}
In addition, we impose the following parametric form on $q(\{\Matrix{\theta}^\lambda \})$ and $q(\mathcal{M})$:
\begin{align}
    q(\{\Matrix{\theta}^\lambda \})q(\mathcal{M})  = \Gaussian \Big(\Matrix{\omega}, \Diag( \exp \{\Matrix{\psi}\} ) \Big).
    \label{eq:diag}
\end{align}
% We further assume that the posteriors of the subspace parameters and embeddings are Gaussian with diagonal covariance. Note that our inference scheme can work with full covariance matrices but the covariance matrices can be quite large, especially for $q(\mathcal{M})$.
% We then maximize the evidence lower bound with respect to $q(\cdot)$:
With the factorization in \eqref{eq:meanfield}, the  variational lower-bound becomes:
\begin{align}
    \mathcal{L}[q] = & \sum_{\lambda=1}^L  \bigg[ \Bigl \langle \ln p(\Matrix{X}^\lambda |\Matrix{z}^\lambda, \Matrix{\theta}^\lambda, \mathcal{M})\Bigr \rangle_q - \KL \left (q(\Matrix{z}^\lambda) || p(\Matrix{z}^\lambda) \right ) \nonumber \\
    &- \KL \left (q(\Matrix{\theta}^\lambda) || p(\Matrix{\theta}^\lambda) \right ) \bigg] - \KL\Bigl (q(\mathcal{M})||p(\mathcal{M})\Bigr). \label{eq:objective}
    % \nonumber \\
    % & - \sum_{\lambda=1}^L \sum_{u=1}^{U^\lambda} \KL \left (q(\Matrix{e}^{\lambda, u}) || p(\Matrix{e}^{\lambda, u}) \right ).
\end{align}
We optimize \eqref{eq:objective} through an \emph{expectation-maximization} procedure where we iteratively re-estimate each of the variational posteriors $q(\{ \Matrix{z}^\lambda \})$ and $q(\{ \Matrix{\theta}^\lambda \})q( \mathcal{M})$ given the current estimate of the other.

In the \emph{expectation} step, we compute $q(\{ \Matrix{z}^\lambda \})$ which maximizes~\eqref{eq:objective} using a modified forward-backward algorithm. Instead of the log-likelihood, we use the expectation of the log-likelihood with respect to the posterior of the the HMM parameters. More details on this can be found in~\cite{ondel2020thesis}.

In the \emph{maximization} step, we compute $q(\{ \Matrix{\theta}^\lambda \})q(\mathcal{M})$ which maximizes~\eqref{eq:objective} 
%given the current estimate of $q(\{ \Matrix{z}^\lambda \})$
. Since this has no closed-form solution, we instead optimize an empirical approximation of \eqref{eq:objective}:

\begin{align}
    \mathcal{L}(\Matrix{\omega}, \Matrix{\psi}) = & \frac{1}{S}\sum_{s=1}^S \bigg\{ \sum_{\lambda=1}^L \bigg[ \Bigl \langle \ln p(\Matrix{X}^\lambda |  \Matrix{z}^\lambda, \Matrix{\theta}_s^\lambda, \mathcal{M}_s) \Bigr \rangle_{q(\Matrix{z}^\lambda)} \nonumber \\
    &- \KL \left (q(\Matrix{\theta}_s^\lambda) || p(\Matrix{\theta}_s^\lambda) \right ) \bigg] - \KL \left (q(\mathcal{M}_s) || p(\mathcal{M}_s)  \right ) \bigg\},
    \label{eq:elbo_samp} \\
    (\{ \Matrix{\theta}_s^\lambda \}, \mathcal{M}_s)  &= \Matrix{\omega} + \exp\{\frac{\Matrix{\psi}}{2}\} \odot \Matrix{\epsilon}_s, \quad  \Matrix{\epsilon}_s  \sim \Gaussian(\Matrix{0}, \Matrix{I}) \label{eq:samp},
\end{align}
where $\odot$ is the element-wise multiplication operation. Equations~\ref{eq:elbo_samp}~-~\ref{eq:samp} are a special case of the so-called re-parameterization trick~\cite{kingma2013auto}. We use stochastic gradient ascent to maximize \eqref{eq:elbo_samp} with respect to $\Matrix{\omega}$ and $\Matrix{\psi}$.

The first term in~\eqref{eq:elbo_samp} is a sum over terms computed separately for each Gaussian component $c=(\lambda, u, i, j)$---$j$th Gaussian component of the $i$th state of unit $u$ in language $\lambda$:
\begin{align}
    \Bigl \langle \ln p(\Matrix{X}^\lambda|\Matrix{\theta}_s^\lambda, \Matrix{z}^\lambda,c) \Bigr \rangle _{q(\Matrix{z}^\lambda)} = & \frac{N_c}{2}\bigl(2\pi^c_s + \ln|\Matrix{\Lambda}^c_s| - \Matrix{\mu}^{c \top}_s \Matrix{\Lambda}^{c}_s \Matrix{\mu}^c_s \bigr) \nonumber \\
    &+ \Matrix{\phi}_c ^\top \Matrix{\Lambda}^{c}_s \Matrix{\mu}^c_s - \frac{\Tr \{\Matrix{\Phi}_c \Matrix{\Lambda}^c_s \}}{2} + \const.
    \label{eq:per_gauss}
\end{align}
For each component, the mean $\Matrix{\mu}_s^c$, precision matrix $\Matrix{\Lambda}^c_s = (\Matrix{\Sigma}_s^c)^{-1}$ and mixing weight ${\pi}_s^c$ are obtained from a sample $(\mathcal{M}_s, \Matrix{\theta}_s^\lambda)$ using equations \ref{eq:f_definition_pi}-\ref{eq:f_definition_mu}.
$\gamma_{nc}$ are the Gaussian responsibilities for each time frame $n$ computed in the \emph{expectation} step, $N_c$ = $\sum_n \gamma_{nc}$, $\Matrix{\phi}_c$ = $\sum_n \gamma_{nc} \Matrix{x}_n$ and $\Matrix{\Phi}_c$ = $\sum_n \gamma_{nc} \Matrix{x}_n \Matrix{x}_n^\top$ are the zeroth, first and second order sufficient statistics respectively for Gaussian component $c$. Note that~\eqref{eq:per_gauss} is a differentiable function of $\Matrix{\omega}$ and $\Matrix{\psi}$.% by the definition of $f(\cdot)$ in~\eqref{eq:f_definition_pi}-\eqref{eq:f_definition_mu}.

Overall, the training of our AUD system comprises two stages. First, we infer a set of variational posteriors $q_0(\{ \Matrix{z}^\lambda \})$, $q_0 (\{ \Matrix{\theta}^\lambda \})$ and $q_0(\mathcal{M})$ on transcribed source languages where $\lambda \in \{1, \dots, L \}$. At this stage, the phone-loop of the AUD model is replaced with a \emph{forced alignment} graph. Then, on the target language $t$, we infer new variational posteriors $q_1(\Matrix{z}^t)$, $q_1 (\Matrix{\theta}^t)$ using $q_0(\mathcal{M})$ to compute the \emph{expectation} and \emph{maximization} steps. Note that $q_0(\mathcal{M})$ is not updated during this stage, but transferred as is from the source languages.% Finally, we use a modified Viterbi algorithm \cite{ondel2020thesis} to decode the units on the target data. 

Finally, the output of our AUD system is obtained from a modified Viterbi algorithm~\cite{ondel2020thesis} which uses the expectation of the log-likelihoods with respect to $q_1(\Matrix{\theta}^t)q_0(\mathcal{M})$ instead of point estimates.

% Overall, the training comprises two parts:
% \begin{enumerate}
%     \item Estimation of $\mathcal{M}$ from transcribed source languages, and
%     \item Learning language and unit embeddings ($\Matrix{\alpha}^\lambda$, $\Matrix{e}^{\lambda,u}$) in the target languages with fixed $\mathcal{M}$.
% \end{enumerate}
% % \Bolaji{In a sense, if we think of $[\Matrix{M}_0, \dots, \Matrix{M}_K]$ as a set of language templates, our task is to learn a weighted sum of these templates for the new language. /*Probably add this to the training section instead*/}
% % In the first part, we learn a set of language templates $\mathcal{M}$. In the second, we learn the `right' weighting factors $\Matrix{alpha}$ required to construct our target language from these along with the embeddings $\Matrix{e}$ required to construct units within that language.
% We optimize the same objective function for both, but make slight modifications to account for the different sources of knowledge available. In the first part, we do the forward-backward with a \textit{forced alignment} graph to take advantage of the phonetic transcriptions. In the second part, we use a pure phone loop (\textit{decoding graph}) as in~\cite{ondel2016variational}.

\section{Related work}
\label{sec:related}
Our proposed model builds heavily on the generalized subspace model of~\cite{Ondel2019shmm}. Our novelty is that we introduce a hyper-subspace to allow unsupervised adaptation of the subspace itself. In particular, if the language embedding $\Matrix{\alpha}$ is set to $\Matrix{0}$, then the H-SHMM becomes identical to the SHMM. Another related model is the Subspace Gaussian Mixture Model (SGMM) used for ASR acoustic modeling~\cite{POVEY2011404}. While the SGMM incorporates a subspace of means of individual Gaussian components and mixture weights, following the SHMM, our subspace models entire 3-state HMMs including covariance matrices. Moreover, where the SGMM uses maximum likelihood training, we use a variational Bayes framework to infer the posterior distributions of the parameters.

% Our proposed model builds heavily on the generalized subspace model of~\cite{Ondel2019shmm}. Our novelty is that we introduce a hyper-subspace to allow unsupervised adaptation of the subspace itself. In particular, if the language embedding $\Matrix{\alpha}$ is set to $\Matrix{0}$, then the H-SHMM becomes identical to the SHMM. Another related model is the Subspace Gaussian Mixture Model (SGMM) used for ASR acoustic modeling~\cite{POVEY2011404}. While the SGMM incorporates a subspace of means of individual Gaussian components and mixture weights, following the SHMM, we include covariance matrices as well for all three states. Moreover, where the SGMM uses maximum likelihood training, we use a variational Bayes framework to infer the posterior distributions of the parameters.

\section{Experiments}
\label{sec:experiments}
\subsection{Data and features}
We test the performance of our models on the following languages: \begin{enumerate}
    \item Mboshi~\cite{godard2017very}: 4.4 hours with 5130 utterances by 3 speakers.
    \item Yoruba~\cite{gutkin2020yoruba}: 4 hours with 3583 utterances by 36 speakers.
    \item English: from TIMIT~\cite{garofolo1990darpa} excluding the \texttt{sa} utterances. 3.6 hours with 4288 utterances by 536 speakers.
\end{enumerate}
Note that we train and test on the entirety of each corpus since we are doing unsupervised learning. We include English in order to have a control language to facilitate comparison with other baselines.

We use seven transcribed languages for training the hyper-subspace: German, Spanish, French and Polish from Globalphone~\cite{schultz2013globalphone}; and Amharic~\cite{Abate2005amharic}, Swahili~\cite{gelas2012swahili} and Wolof~\cite{gauthier2016wolof} from the ALFFA project~\cite{besacier2015speech}. For each of these, we use only a subset of 1500 utterances which corresponds to 3-5 hours per language.

%We train with 13-dimensional MFCC features along with their first and second order derivatives. 

\subsection{Metrics}
We evaluate the performance of our models with two metrics for phone segmentation and phone clustering. For segmentation, we report the F-score on phone boundary detection with a tolerance of $\pm 20$ milliseconds. For clustering, we report the normalized mutual information (NMI) which is computed from a frame-level confusion matrix of discovered units ($U$) and actual phones ($P$) as:
\begin{align}
    \text{NMI}(P, U) = 200 \times \frac{I(P; U)}{H(P) + H(U)} \%,
\end{align}
where $H(\cdot)$ is the Shannon entropy and $I(P; U)$ is the (un-normalized) mutual information. An NMI of $0$ means that the the discovered acoustic units are completely unrelated to the actual phones, while an NMI of $100$ means that the units have a one-to-one correspondence with the actual phones. Note that the $H(U)$ term in the denominator rewards more compact representations.

\subsection{Experiment setup}
Although the Dirichlet process prior allows us to model an arbitrary number of units, in practice, we set the truncation parameter~\cite{blei2006variational} to 100, and we find that number of discovered units tends to be fewer. We set the dimension of the H-SHMM unit embeddings $|\Matrix{e}^{\lambda,u}| = 100$. We set the dimension of the language embeddings $|\Matrix{\alpha}^{\lambda}|=6$. We use 5 samples for the re-parametrization trick and train with Adam optimizer~\cite{kingma2014adam} with a learning rate of $5\times10^{-3}$.
% since we have seven source languages.
% for the AUD proper since we have seven source languages, but we also train a model with language embeddings of dimension 2 to facilitate visualization of the learned embeddings.

We report results on five baselines: HMM~\cite{ondel2016variational}, SHMM~\cite{Ondel2019shmm}, VQ-VAE~\cite{van2017neural}, VQ-wav2vec~\cite{Baevski2020vq-wav2vec} and ResDAVEnet~\cite{Harwath2020Learning}. We use 13-dimensional MFCC features along with their first and second derivatives as features for the H-SHMM, HMM, SHMM and VQ-VAE.%We re-implemented the first three ourselves and used pre-trained models for the other two. 

The HMM~\cite{ondel2016variational} and SHMM~\cite{Ondel2019shmm} are the most comparable models to the H-SHMM. We set the truncation parameter to 100 for both. Furthermore, we train the SHMM subspace with the same source languages that we use to train the H-SHMM hyper-subspace and we set the subpace dimension to 100.

We also impemented VQ-VAE~\cite{van2017neural} baselines as they have been shown to learn good discrete representations of speech~\cite{chorowski2019unsupervised}. The most critical choices we made were: (i) having a big encoder (5 BLSTM layers) but weak decoder (feed-forward with one hidden layer) as stronger decoders resulted in better reconstruction error but worse latent representations, (ii) low-dimensional latent space (16-d) to discard irrelevant information, (iii) relatively few units (50 centroids) and down-sampled encoder (by factor of 2) to prevent over-segmentation, and (iv) concatenating a learned speaker-embedding (32-d) to the decoder input to help the encoder focus more on phonetic information. We tune these parameters to maximize the NMI and F-score on English and kept them for the other two languages.

We report results for two other neural discrete representation learning models: VQ-wav2vec ~\cite{Baevski2020vq-wav2vec} and ResDAVEnet-VQ~\cite{Harwath2020Learning}. Note that we cannot replicate these models on our low-resource languages as the former is trained with $960$ hours of Librispeech~\cite{panayotov2015librispeech} data while the latter requires paired captioned images for training~\cite{zhou2014learning}. Therefore, in both cases, we use the authors' own code and pre-trained models to extract discrete representations for the TIMIT utterances and report those results. We tested the various pre-trained models available for each system and report only the best performing ones.

For VQ-wav2vec, we report results for the Gumbel softmax variant since it gave higher NMI and F-score than the K-means variant.

For ResDAVEnet-VQ, we report results on the ``$\{3\} \rightarrow \{2,3\}$" model (ResDAVEnet-VQ-I) with units extracted from layer-2 and the ``$\{2\} \rightarrow \{2,3\}$" model (ResDAVEnet-VQ-II) with units extracted from layer-3 as we found
they had the highest NMI and F-score respectively of the available pre-trained models.

Our H-SHMM, HMM and SHMM code are publicly available~\footnote{{https://github.com/beer-asr/beer/tree/master/recipes/hshmm}}. Our implementation of the VQ-VAE is also public~\footnote{{https://github.com/BUTSpeechFIT/vq-aud}}.

% Our model implementations are all available on Github. Our HMM, SHMM and HSHMM implementations are in the Beer toolkit~\footnote{{https://github.com/beer-asr/beer/tree/master/recipes/hshmm}}. Our implementation of the VQ-VAE is also on Github~\footnote{{https://github.com/BUTSpeechFIT/vq-aud}}.% The pre-trained VQ-wav2vec and ResDAVEnet-VQ code and models are on the respective authors' own repositories.

\subsection{Experiment results}
\begin{table}[t]
    \centering
    \caption{Acoustic unit discovery results.}
    \resizebox{0.43\textwidth}{!}{%
    \begin{tabular}{llcc}
    \toprule
         Corpus & System & NMI & F-score  \\ \midrule
         English & ResDAVEnet-VQ-I& 35.93 & 54.19 \\
         & ResDAVEnet-VQ-II& 34.39 & 64.36 \\
         & VQ-wav2vec & 35.20 & 26.84 \\
         & VQ-VAE & 32.03 $\pm$ 0.30 & 59.05 $\pm$ 0.34 \\

         & HMM & 35.91 $\pm$ 0.27 & 63.86 $\pm$ 0.95 \\
         & SHMM & 39.17  $\pm$  0.16 &  74.65  $\pm$  0.60 \\
         & SHMM+finetune & 37.83 $\pm$ 0.25 & 72.20 $\pm$ 0.65 \\
          & SHMM+300d & 39.62 $\pm$ 0.20 & 73.62 $\pm$ 0.84 \\
         & H-SHMM (ours) & \textbf{40.04} $\pm$ 0.51 & \textbf{76.60} $\pm$ 0.54 \\
         \midrule
         
         Mboshi & VQ-VAE & 31.27 $\pm$ 0.26 & 39.19 $\pm$ 0.71 \\
         & HMM & 35.85 $\pm$ 0.62 &  47.92 $\pm$ 1.56 \\
         & SHMM & 38.38 $\pm$ 0.97 & \textbf{59.50} $\pm$ 0.78 \\
         & SHMM+finetune & 36.09 $\pm$ 0.49 & 53.06 $\pm$ 1.06 \\
         & SHMM+300d & 37.51 $\pm$ 0.45 & 53.71 $\pm$ 1.41 \\
         & H-SHMM (ours) & \textbf{41.07} $\pm$ 1.09 & 59.15 $\pm$ 1.51 \\ \midrule
         
         Yoruba & VQ-VAE & 29.90 $\pm$ 0.40 & 37.52 $\pm$ 0.79 \\
         & HMM & 36.38 $\pm$ 0.22 & 54.47 $\pm$ 0.64 \\
         & SHMM & 38.99 $\pm$ 0.08 & 64.46 $\pm$ 0.51 \\
         & SHMM+finetune & 36.97 $\pm$ 0.38 & 58.59 $\pm$ 0.34 \\
         & SHMM+300d & 39.08 $\pm$ 0.13 & 61.09 $\pm$ 1.01 \\
         & H-SHMM (ours) &  \textbf{40.06} $\pm$ 0.11 & \textbf{66.95} $\pm$ 0.36 \\
        \bottomrule
    \end{tabular}
    }
    \label{tab:aud_results}
\end{table}

We train each system with 5 random initializations and report the means and standard deviations of the results in Table~\ref{tab:aud_results}. The SHMM and H-SHMM subspace and hyper-subspace respectively are trained once, and only the AUD is repeated 5 times. Since we use pre-trained VQ-wav2vec and ResDAVEnet-VQ models, we only run them once per language. From the results, we take the SHMM as our main baseline since it outperforms the other baselines on all metrics. Moreover, it provides the best direct comparison as it is the most structurally similar baseline to our proposed model.
% This is not unexpected as it is the only one that is trained specifically for phonetic discovery. 
% only outperforms the other baselines on all metrics, it is also the most structurally similar to our proposed model. This is not unexpected as it is the only one that is trained specifically for phonetic discovery.

We achieve significant NMI improvements over the SHMM with the H-SHMM. %\Bolaji{Even if we subtract the standard deviations from the H-SHMM averages while adding the standard deviations to the SHMM averages, the H-SHMM still performs better.}
We also get similar F-score improvements in English and Yoruba, but get a slightly worse average F-score on Mboshi.

The novelty of the H-SHMM is that we introduce a way of adapting the subspace to the target language. We tested whether simply fine-tuning the SHMM \textit{subspace} parameters on the target language would achieve the same results. The result of fine-tuning (SHMM+finetune) is not just worse than H-SHMM, it is in fact worse than SHMM. This is intuitive because fine-tuning the subspace parameters relaxes the constraint on the HMM parameters.% that makes the SHMM good in the first place.

Another difference is that the H-SHMM has more transferred parameters than the SHMM. Therefore, we experimented with increasing the number of transferred SHMM parameters by changing the subspace dimension to 300 (SHMM+300d). The results show no significant benefit over the SHMM with dimension 100.
% to see if this approaches the performance of the H-SHMM

To visualize the learned language embeddings, we train an H-SHMM with $|\Matrix{\alpha}^\lambda|=2$. For this experiment, we split each corpus into four subsets, so that the model \textit{a priori} treats each subset as a different language. After inference, we find that $\Matrix{\alpha}^\lambda$ of different subsets of the same language converge with small within-language variance for source languages and higher variance for target languages, so the model is able to cluster subsets that come from the same language without being told. The images can be found at \emph{https://www.fit.vutbr.cz/{\texttildelow}iyusuf/hshmm.html}.

% with splitting each language into four non-overlapping subsets, and training an H-SHMM with $|\Matrix{\alpha}^\lambda|=2$ on these. We find that the embeddings of different subsets of the same language converge with small within-language variance for source languages and higher variance for target languages. The images can be found at \emph{http://merlin.fit.vutbr.cz/iondel/hshmm/}.

% We find that the HMM baseline is significantly better than the VQ-VAE both in terms of NMI and segmentation despite the latter also incorporating speaker information. This suggests that the structural prior inherent in the 3-state HMM-imposing smooth trajectories-is already quite good for modeling phones, although down-sampling also imposes a similar prior on the VQ-VAE. On English VQ-wav2vec meanwhile achieves only slightly worse NMI than the HMM, but the F-score is much worse since it uses thousands of units, and therefore over-segments the data. The ResDAVEnet models achieve comparable NMI and F-score depending on which configuration is used. However, the SHMM performs considerably better than either of these baselines in all metrics, so we consider it our main baseline.
% \subsection{Language embeddings}

\section{Conclusion}
\label{sec:conclusion}
In this paper, we have proposed a hierarchical subspace model for unsupervised phonetic discovery in which a phonetic subspace constrains the parameters and ensures that the learned parameters define a plausible phone. Similarly, a hyper-subspace constrains the parameters of the subspace itself. We have shown that the proposed model outperforms the non-hierarchical baseline as well as other neural network-based AUD models.

Going forward, we hope to explore better generative models than the HMM, as well as different subspace hierarchies, such as having different subspaces for various phone classes or speaker subspaces.

% \section{Acknowledgment}
% \label{sec:acknowledgement}
% \input{inputfiles/acknowledgment}

\bibliographystyle{IEEEbib}
\bibliography{refs}

\begin{thebibliography}{10}

\bibitem{dupoux2018cognitive}
Emmanuel Dupoux,
\newblock ``Cognitive science in the era of artificial intelligence: A roadmap
  for reverse-engineering the infant language-learner,''
\newblock {\em Cognition}, vol. 173, pp. 43--59, 2018.

\bibitem{versteegh2015zero}
Maarten Versteegh et~al.,
\newblock ``{The Zero Resource Speech Challenge 2015},''
\newblock in {\em Sixteenth annual conference of the international speech
  communication association}, 2015.

\bibitem{dunbar2017zero}
Ewan Dunbar et~al.,
\newblock ``{The Zero Resource Speech Challenge 2017},''
\newblock in {\em 2017 IEEE Automatic Speech Recognition and Understanding
  Workshop (ASRU)}. IEEE, 2017, pp. 323--330.

\bibitem{dunbar2019zero}
Ewan Dunbar et~al.,
\newblock ``{The Zero Resource Speech Challenge 2019: TTS Without T},''
\newblock in {\em Interspeech}, 2019, pp. 1088--1092.

\bibitem{chorowski2019unsupervised}
Jan Chorowski, Ron~J Weiss, Samy Bengio, and A{\"a}ron van~den Oord,
\newblock ``{Unsupervised speech representation learning using wavenet
  autoencoders},''
\newblock {\em IEEE/ACM transactions on audio, speech, and language
  processing}, vol. 27, no. 12, pp. 2041--2053, 2019.

\bibitem{Baevski2020vq-wav2vec}
Alexei Baevski, Steffen Schneider, and Michael Auli,
\newblock ``{vq-wav2vec: Self-Supervised Learning of Discrete Speech
  Representations},''
\newblock in {\em International Conference on Learning Representations}, 2020.

\bibitem{van2017neural}
Aaron Van Den~Oord, Oriol Vinyals, et~al.,
\newblock ``Neural discrete representation learning,''
\newblock in {\em Advances in Neural Information Processing Systems}, 2017, pp.
  6306--6315.

\bibitem{lee2012nonparametric}
Chia-ying Lee and James Glass,
\newblock ``{A nonparametric Bayesian approach to acoustic model discovery},''
\newblock in {\em Proceedings of the 50th Annual Meeting of the Association for
  Computational Linguistics: Long Papers-Volume 1}. Association for
  Computational Linguistics, 2012, pp. 40--49.

\bibitem{ondel2016variational}
Lucas Ondel, Luk{\'a}{\v{s}} Burget, and Jan {\v{C}}ernock{\`y},
\newblock ``{Variational inference for acoustic unit discovery},''
\newblock {\em Procedia Computer Science}, vol. 81, pp. 80--86, 2016.

\bibitem{chen2017multilingual}
Hongjie Chen, Cheung-Chi Leung, Lei Xie, Bin Ma, and Haizhou Li,
\newblock ``Multilingual bottle-neck feature learning from untranscribed
  speech,''
\newblock in {\em 2017 IEEE Automatic Speech Recognition and Understanding
  Workshop (ASRU)}. IEEE, 2017, pp. 727--733.

\bibitem{glarner2018full}
Thomas Glarner, Patrick Hanebrink, Janek Ebbers, and Reinhold Haeb-Umbach,
\newblock ``{Full Bayesian Hidden Markov Model Variational Autoencoder for
  Acoustic Unit Discovery},''
\newblock in {\em Interspeech}, 2018, pp. 2688--2692.

\bibitem{Ondel2019shmm}
Lucas Ondel, Hari~Krishna Vydana, Lukáš Burget, and Jan Černocký,
\newblock ``{Bayesian Subspace Hidden Markov Model for Acoustic Unit
  Discovery},''
\newblock in {\em Interspeech}, 2019, pp. 261--265.

\bibitem{Teh2010a}
Y.~W. Teh,
\newblock ``{D}irichlet processes,''
\newblock in {\em Encyclopedia of Machine Learning}. Springer, 2010.

\bibitem{ondel2020thesis}
Lucas Ondel,
\newblock {\em Discovering Acoustic Units from Speech: A Bayesian Approach},
\newblock Ph.D. thesis, Brno University of Technology, Faculty of Information
  Technology, to appear.

\bibitem{ondel2018bayesian}
Lucas Ondel et~al.,
\newblock ``{Bayesian models for unit discovery on a very low resource
  language},''
\newblock in {\em 2018 IEEE International Conference on Acoustics, Speech and
  Signal Processing (ICASSP)}. IEEE, 2018, pp. 5939--5943.

\bibitem{ondel2017bayesian}
Lucas Ondel, Luka{\v{s}} Burget, Jan {\v{C}}ernock{\`y}, and Santosh Kesiraju,
\newblock ``{Bayesian phonotactic language model for acoustic unit
  discovery},''
\newblock in {\em 2017 IEEE International Conference on Acoustics, Speech and
  Signal Processing (ICASSP)}. IEEE, 2017, pp. 5750--5754.

\bibitem{kingma2013auto}
Diederik~P Kingma and Max Welling,
\newblock ``{Auto-encoding Variational Bayes},''
\newblock in {\em ICLR}, 2014.

\bibitem{POVEY2011404}
Daniel Povey et~al.,
\newblock ``The subspace gaussian mixture model—a structured model for speech
  recognition,''
\newblock {\em Computer Speech \& Language}, vol. 25, no. 2, pp. 404 -- 439,
  2011,
\newblock Language and speech issues in the engineering of companionable
  dialogue systems.

\bibitem{godard2017very}
Pierre Godard et~al.,
\newblock ``A very low resource language speech corpus for computational
  language documentation experiments,''
\newblock {\em arXiv preprint arXiv:1710.03501}, 2017.

\bibitem{gutkin2020yoruba}
Alexander Gutkin, I{\c{s}}{\i}n Demir{\c{s}}ahin, Oddur Kjartansson, Clara
  Rivera, and K\d{\'o}lá Túb\d{\`o}sún,
\newblock ``{Developing an Open-Source Corpus of Yoruba Speech},''
\newblock in {\em Interspeech}, Shanghai, China, 2020.

\bibitem{garofolo1990darpa}
J~Garofolo, L~Lamel, W~Fisher, J~Fiscus, D~Pallet, and N~Dahlgren,
\newblock ``{The DARPA TIMIT acoustic-phonetic continuous speech corpus CDROM.
  NTIS order number PB91-505065},'' 1990.

\bibitem{schultz2013globalphone}
Tanja Schultz, Ngoc~Thang Vu, and Tim Schlippe,
\newblock ``Globalphone: A multilingual text \& speech database in 20
  languages,''
\newblock in {\em 2013 IEEE International Conference on Acoustics, Speech and
  Signal Processing}. IEEE, 2013, pp. 8126--8130.

\bibitem{Abate2005amharic}
Solomon~Teferra Abate, Wolfgang Menzel, and Bairu Tafila,
\newblock ``{An Amharic Speech Corpus for Large Vocabulary Continuous Speech
  Recognition},''
\newblock in {\em INTERSPEECH-2005}, 2005.

\bibitem{gelas2012swahili}
Hadrien Gelas, Laurent Besacier, and Francois Pellegrino,
\newblock ``{D}evelopments of {S}wahili resources for an automatic speech
  recognition system,''
\newblock in {\em {SLTU} - {W}orkshop on {S}poken {L}anguage {T}echnologies for
  {U}nder-{R}esourced {L}anguages}, Cape-Town, Afrique Du Sud, 2012.

\bibitem{gauthier2016wolof}
Elodie Gauthier, Laurent Besacier, Sylvie Voisin, Michael Melese, and
  Uriel~Pascal Elingui,
\newblock ``{Collecting Resources in Sub-Saharan African Languages for
  Automatic Speech Recognition: a Case Study of Wolof},''
\newblock {\em LREC}, 2016.

\bibitem{besacier2015speech}
Laurent Besacier et~al.,
\newblock ``Speech technologies for african languages: Example of a
  multilingual calculator for education,''
\newblock in {\em Interspeech}, 2015.

\bibitem{blei2006variational}
David~M Blei, Michael~I Jordan, et~al.,
\newblock ``{Variational inference for Dirichlet process mixtures},''
\newblock {\em Bayesian analysis}, vol. 1, no. 1, pp. 121--143, 2006.

\bibitem{kingma2014adam}
Diederik~P Kingma and Jimmy Ba,
\newblock ``Adam: A method for stochastic optimization,''
\newblock {\em arXiv preprint arXiv:1412.6980}, 2014.

\bibitem{Harwath2020Learning}
David Harwath, Wei-Ning Hsu, and James Glass,
\newblock ``{Learning Hierarchical Discrete Linguistic Units from
  Visually-Grounded Speech},''
\newblock in {\em International Conference on Learning Representations}, 2020.

\bibitem{panayotov2015librispeech}
Vassil Panayotov, Guoguo Chen, Daniel Povey, and Sanjeev Khudanpur,
\newblock ``{Librispeech: an ASR corpus based on public domain audio books},''
\newblock in {\em 2015 IEEE International Conference on Acoustics, Speech and
  Signal Processing (ICASSP)}. IEEE, 2015, pp. 5206--5210.

\bibitem{zhou2014learning}
Bolei Zhou, Agata Lapedriza, Jianxiong Xiao, Antonio Torralba, and Aude Oliva,
\newblock ``Learning deep features for scene recognition using places
  database,''
\newblock in {\em Advances in neural information processing systems}, 2014, pp.
  487--495.

\end{thebibliography}

\end{document}